\documentclass[reprint, aip, amsmath,amssymb, superscriptaddress, jcp, citeautoscript]{revtex4-2}
\usepackage{graphicx}% Include figure files
\usepackage{dcolumn}% Align table columns on decimal point
\usepackage{bm}% bold math
\usepackage{braket}
\usepackage{verbatim}
\usepackage{lipsum}
\usepackage{enumitem}
\usepackage{makecell}
\usepackage{relsize}
\usepackage{hyperref}
\usepackage{xr}
\usepackage[dvipsnames]{xcolor}
\usepackage{bbold}
\usepackage{cases}
\usepackage{upgreek}
\usepackage[normalem]{ulem}
\usepackage{textgreek}
\usepackage{gensymb}
\usepackage{amssymb}

\newcommand{\beq}{\begin{equation}}
\newcommand{\eeq}{\end{equation}}
\newcommand{\beqarray}{\begin{eqnarray}}
\newcommand{\eeqarray}{\end{eqnarray}}
\newcommand{\Sec}[1]{Sec.~\ref{#1}} % Sec. label

\begin{document}

\title{ 
Systematically improvable semiclassical clustering for accurate and efficient quantum dynamics in large spin ensembles
}

\author{Pranay Venkatesh}
\affiliation{Department of Chemistry, University of Colorado Boulder, Boulder, Colorado 80309, USA\looseness=-1}
\author{Andr\'es Montoya-Castillo}
\homepage{Andres.MontoyaCastillo@colorado.edu}
\affiliation{Department of Chemistry, University of Colorado Boulder, Boulder, Colorado 80309, USA\looseness=-1}
\date{\today}

\begin{abstract}
Predicting the decoherence of spin qubits in realistic environments is key to identifying the interactions that limit their coherence---a prerequisite for designing better quantum technologies. While accurate methods for spin dynamics are limited to relatively small systems, the dissipative spin environments in real experiments typically involve orders of magnitude more spins. Semiclassical methods can reach these sizes, but their classical treatment of the bath cannot capture the sensor-bath entanglement that generates coherence revivals, especially under the application of refocusing pulses. Here, we resolve this tension. We introduce cluster spin-LSC, a method that clusters the sensor together with a few strongly coupled bath spins, treating them quantum mechanically in an otherwise classical spin bath. We show that our method is applicable to systems containing arbitrary spin species under dynamical decoupling pulse sequences. Our method converges systematically to the exact dynamics with cluster size, providing an internal consistency check when no exact benchmarks exist. We demonstrate that cluster spin-LSC recovers numerically exact coherence and population dynamics for a nitrogen-vacancy center and a molecular qubit, reducing day-long multi-GPU simulations to only minutes on a CPU. We further extend our method to systems with large spin baths that lie beyond the reach of exact methods. Our approach thus offers a scalable and systematically improvable route to accurately simulate the decoherence dynamics of sensors in realistic spin environments.
\end{abstract}

\maketitle

\section{Introduction}

Spins in solid-state defects, semiconductor quantum dots, and molecules are promising platforms for quantum sensors~\cite{DegenRevModPhys2017, maze2008nanoscale, dolde2011electric, Lovchinsky2016}, qubits, and quantum memories~\cite{stanwix2010coherence, gangloff2019quantum, Weiss2025} owing to their long coherence times~\cite{zadrozny2015millisecond, awschalom2018quantum}. To guide their design, one needs to know which interactions cause decoherence. Experiments have shown how coherence times depend on the applied magnetic field, temperature, and chemical structure of the host~\cite{Jackson2022JPCC, martinez2023impact, mullin2023quantum}, but these trends do not identify the microscopic mechanisms underlying decoherence, making it difficult to guide the design of near-term quantum technologies. Achieving this requires methods that can accurately simulate the dynamics of many-spin Hamiltonians over long timescales and at experimentally relevant system sizes.

Recently, we developed the spin bath-truncated matrix product state (SB-tMPS) method~\cite{Li2026}, which provides numerically exact coherence and population dynamics of interacting spin-bath Hamiltonians under arbitrary pulse sequences. However, its cost grows with the entanglement between the sensor and its bath: GPU-accelerated simulations require hours for $\sim$100 spins and are limited to $\sim$200 bath spins in the weakly interacting limit. Molecular qubits in frozen solutions and electron spins in quantum dots have environments of hundreds to tens of thousands of nuclear spins, making them inaccessible to numerically exact simulations.

Semiclassical approaches offer a more scalable alternative. At the temperatures of experiments, which for spins are effectively infinite, semiclassical treatments start to become physically meaningful. Schulten-Wolynes theory replaces the nuclear spins hyperfine-coupled to an electron with a static, classical Overhauser field~\cite{Schulten1978}, a picture later adopted for central spin models~\cite{Merkulov2002} and extended to realistic spin systems~\cite{Manolopoulos2013, lewis2014asymmetric, Stanek2014, lewis2016efficient, Fauseweh2017, Fay2017, Fay2019}. These methods describe radical pairs and quantum dots accurately. However, quantum sensing leverages \textit{entanglement} between the sensor and individual bath spins, which a classical description of the bath struggles to capture. Further, dynamical decoupling pulse sequences extend coherence lifetimes and allow for entanglement between the sensor and its environment to build and spread, thereby intensifying this problem.

One potential way to overcome this limitation is through clustering, where one treats subsets of spins quantum mechanically and their surroundings semiclassically. Clustering extended the applicability of the truncated Wigner approximation (CTWA)~\cite{Wurtz2018} and spin dynamic mean-field theory (CspinDMFT)~\cite{grasser2023understanding, bieniek2026dynamical}, enabling these methods to accurately capture the physics of localization in disordered spin chains and spin diffusion in dipolar ensembles. In quantum sensing, cluster-based approaches rely predominantly on factorization~\cite{Krogmeier2026} and underlie the cluster correlation expansion (CCE)~\cite{Yang2008CCE1, Yang2009CCE2}, which factorizes the sensor coherence as a product of irreducible contributions from small clusters of bath spins. Building these irreducible factors requires dividing subcluster contributions, which engenders numerical instabilities when those contributions approach zero. The resulting dynamics can diverge and become unphysical, and increasing the order of the expansion fails to fix the problem or result in systematically improvable dynamics~\cite{Li2026, Krogmeier2026, ryan2026challenge}.

Here, we develop and demonstrate the accuracy and scalability of our cluster spin-LSC (CSLSC) method, which embeds a small quantum mechanical cluster in a classical spin bath within the linearized semiclassical initial value representation (LSC-IVR) framework. Our method exploits the observation that most bath spins dephase the sensor through Ising-type interactions, which are well described under classical approximation~\cite{witzel2014converting}. Coherence revivals arise from a small number of nuclei with appreciable anisotropic hyperfine couplings or strong quadrupolar interactions, whose entanglement with the sensor requires a quantum mechanical treatment. We employ the spin-mapping framework~\cite{runeson2019, runeson2020} to map the sensor and these nuclei onto a single multilevel system and treat the remaining bath spins classically. Our CSLSC handles nuclei of arbitrary spin, treats zero-field splittings and quadrupolar interactions exactly, and holds for arbitrary dynamical decoupling pulse sequences. Its results converge systematically with the number of bath spins in the cluster, providing an implicit confirmation of accuracy when no numerically exact benchmark exists. We demonstrate the applicability of our CSLSC on three representative spin qubit problems of increasing complexity, with the last one lying squarely beyond the applicability of other accurate methods: an NV center in diamond, an erbium molecular qubit, and an InAs quantum dot system.

\section{Spin Hamiltonian}
\label{sec:hamiltonian}

We focus on general Hamiltonians describing a central spin (sensor) coupled to an interacting spin bath. We organize terms such that the Hamiltonian splits into interactions that standard semiclassical treatments capture faithfully and those that strain existing methods. The total Hamiltonian,
\begin{align}
 \label{eq:ham}
    H_{\rm T} = H_{\rm s} + H_{\rm b} + H_{\rm s-b} + H_{\rm I},
\end{align}
contains both sensor (\textit{s}) and bath (\textit{b}) contributions,
\begin{align}
    H_{\rm s} &= \omega_s S_0^z + \sum_{k,l\in\{x,y,z\}}S_0^k \mathbf{D}^{kl} S_0^l \label{eq:Hs}, \\
     H_{\rm b} &= \sum_{i=1}^K\omega_iI_i^z + \sum_{i=1}^K \sum_{k,l\in\{x,y,z\}}I_i^k \mathbf{P}_i^{kl} I_i^l \label{eq:Hb},
\end{align}
where $K$ is the total number of bath spins considered. Here, $S_0^z$ and $I_i^z$ are the $z$ spin operators for the central and $i$th bath spin, respectively, and $\omega_s = \gamma_s B_z$ ($\omega_i = \gamma_i B_z$) is the central ($i$th bath) spin's frequency, where $\gamma_s$ ($\gamma_i$) is the gyromagnetic ratio of the central ($i$th bath) spin and $B_z$ is the magnitude of the magnetic field applied in the $z$ direction. The second terms in Eqs.~\ref{eq:Hs} and \ref{eq:Hb}, respectively, correspond to the zero-field splitting for an electronic spin and the quadrupole interaction for nuclear spins with $I>1/2$. The sensor-bath coupling,
\begin{align}
\label{eq:sbham}
     H_{\rm s-b} = \sum_{i=1}^K \sum_{k,l\in\{x,y,z\}} S_0^{k} \mathbf{A}_{i}^{kl} I_i^{l} = H_{\rm s-b}^{\rm sec} + H_{\rm s-b}^{\rm ps} + H_{\rm s-b}^{\rm ns},
\end{align}
involves the coupling tensor, $\mathbf{A}_i$, between the sensor and the $i$th bath spin, which we decompose into secular (sec), pseudosecular (ps), and nonsecular (ns) contributions,
\begin{align}
    H_{\rm s-b}^{\rm sec} &= \sum_{i=1}^K S_0^z \mathbf{A}_i^{zz}  I_i^z, \label{eq:Hsec}\\
    H_{\rm s-b}^{\rm ps} &= \sum_{i=1}^K S_0^z \left(\mathbf{A}_i^{zx} I_i^x + \mathbf{A}_i^{zy} I_i^y \right), \label{eq:Hps}\\
    H_{\rm s-b}^{\rm ns} &= \sum_{i=1}^K \sum_{k\in\{x,y\}} \sum_{l\in\{x,y,z\}} S_0^k \mathbf{A}_i^{kl} I_i^l. \label{eq:Hns}
\end{align}
This decomposition results in a nested hierarchy of regimes. The Ising limit~\cite{yao2006theory, witzel2014converting, jeschke2023nuclear} retains only the secular sensor-bath terms (which contribute to pure dephasing), the full pure-dephasing limit~\cite{Canarie2020JPCL, tarkanyi2026understanding} adds the pseudosecular terms, and the full decoherence limit~\cite{Onizhuk2021PRXQ, tarkanyi2026understanding} further adds the nonsecular terms. We parameterize all couplings from standard geometries of the physical systems (see SI~Sec.~I). The intra-bath contribution,
\begin{align}
\label{eq:HI}
    H_{\rm I} = \sum_{i<j}\sum_{k,l\in\{x,y,z\}} I_i^k \mathbf{J}_{i,j}^{kl}I_j^l,
\end{align}
contains interparticle interactions, where $\mathbf{J}_{i,j}$ denotes the coupling tensor connecting bath spins $i$ and $j$. These interactions are important as they modulate the experimentally observed decay of coherence dynamics under applied pulse sequences~\cite{Zhao2012, Maze2008PRB, Szankowski2017JP}. Under an $n$-pulse Carr-Purcell-Meiboom-Gill (CPMG-$n$)~\cite{carr1954effects, meiboom1958modified} sequence, one applies $n$ instantaneous $\pi$ pulses, $U = \exp(-i\pi\sigma^y_{01}/2)$ such that the total propagator is $e^{-iH_{\rm T}\tau}(U e^{-2iH_{\rm T}\tau})^{n-1}U e^{-iH_{\rm T}\tau}$, where $\tau = t/2n$. Spin echo thus corresponds to $n=1$, while free induction decay (FID) corresponds to $n=0$.

For quantum sensing with a single sensor, our observables of interest are the matrix elements of the reduced density matrix of the sensor spin, $\rho_{\rm s}(t) = \mathrm{Tr}_{\rm B}[\rho(t)]$, where $\rho(t)$ is the complete density matrix of the sensor and the bath. Taking $\{\ket{0}, \ket{1}\}$ as the two sensor levels that define the qubit, the diagonal elements of $\rho_{\rm s}(t)$ yield the sensor populations, while its off-diagonal elements yield the coherence. We thus monitor the coherence and population dynamics,
\begin{align}
    C(t) &= \frac{\lvert\bra{0}\rho_{\rm s}(t)\ket{1}\rvert}{\lvert\bra{0}\rho_{\rm s}(0)\ket{1}\rvert}, \label{eq:coh}\\
    P_1(t) &= \bra{1}\rho_{\rm s}(t)\ket{1}. \label{eq:pop}
\end{align}

We take the sensor and bath to be initially uncorrelated, $\rho(0) = \rho_s\otimes\rho_{b}$, consistent with sensing protocols. We prepare the sensor in a pure state, $\ket{\psi_0} = (\ket{0}+\ket{1})/\sqrt{2}$ and take the bath spins to be at infinite temperature $\rho_{b} \propto \mathbb{1}$, since the spin energies ($\lesssim 100$~MHz) lie far below the thermal energy $k_B T$ at experimental temperatures~\cite{Krogmeier2026, abragam1961principles}. 

In the Ising limit, the sensor couples to the bath only through $H_{\rm s-b}^{\rm sec}$, which commutes with $S_0^z$. The sensor spin then experiences a longitudinal field from $\sum_{i} \mathbf{A}_i^{zz} I_i^z$. In the absence of intra-bath interactions, each $I_i^z$ is conserved and, hence, the coherence of an infinite-temperature bath factorizes into contributions from individual bath spins~\cite{cucchietti2005decoherence}. The sensor's coherence dynamics depends only on the statistics of the conserved $I_i^z$, and phase-space methods that sample these from their quantum distributions reproduce it exactly~\cite{schachenmayer2015, Zhu2019}. Semiclassical treatments of central spin problems have exploited this regime for decades, from the frozen Overhauser field picture of radical pairs and quantum dots~\cite{Schulten1978, Merkulov2002, Fay2019, Fay2017} to more recent spin mapping approaches~\cite{lewis2014asymmetric, runeson2020}. Semiclassics thus offers an accurate and scalable description of the bath spins that contribute to secular sensor-bath coupling. 

Pseudosecular and quadrupolar interactions lie beyond the Ising limit, since neither conserves $I_i^z$. The pseudosecular coupling drives transverse nuclear spin dynamics through the $\mathbf{A}_i^{zx}$ and $\mathbf{A}_i^{zy}$ terms, and a quadrupolar tensor can mix the $I_i^z$ eigenstates for nuclei with $I>1/2$. In this limit, the sensor coherence no longer dephases due to a static Overhauser field, and instead experiences a varying field induced by oscillating $I_i^z$ components. The coherence decay in this limit is no longer monotonic and exhibits revivals under both free induction decay and pulsed measurements. These revivals are the signatures underlying electron spin echo envelope modulation (ESEEM) spectroscopy~\cite{Rowan1965, dikanov, schweiger2001principles}. Since the amplitude of the revivals grows with the strength of the pseudosecular or quadrupolar coupling relative to the nuclear Zeeman frequency~\cite{dikanov}, these coherence revivals are often dominated by a few nuclei that are strongly coupled to the sensor or have appreciable quadrupolar interactions~\cite{Seo2016natcomm}.

Including nonsecular terms pushes the dynamics beyond the pure-dephasing limit since $S_0^z$ is no longer conserved, thus causing population relaxation of the sensor spin. At low fields, population relaxation proceeds on timescales comparable to those of dephasing and contributes to coherence decay~\cite{Krogmeier2026}. In the full decoherence limit, one needs both $C(t)$ and $P(t)$ to characterize decoherence dynamics completely~\cite{ficheux2018dynamics}.

\section{Cluster spin-LSC Method}

Here, we introduce our cluster spin-LSC (CSLSC) method. It simulates the dynamics of large spin systems across all three regimes by evolving classical trajectories while retaining quantum mechanical correlations. We adopt the spin-mapping approach~\cite{runeson2019, runeson2020}, which eliminates the unphysical leakage of zero-point energy in other phase-space mapping approaches. However, direct application of the spin-mapping to individual spins sacrifices inter-spin correlation. To remedy this shortcoming and capture the sensor-bath entanglement that underlies ESEEM-like coherence revivals, we cluster the sensor together with the bath spins that contribute most to pseudosecular coupling terms. Since the cost of evaluating spins outside the cluster grows with the connectedness of the Hamiltonian, we exploit the sparsity of sensor-bath and intra-bath couplings to lower the computational cost.

\subsection{Phase-space representation}
\label{subsec:phase-space}

The linearized semiclassical (LSC) approximation expresses observables as averages over classical trajectories in phase space~\cite{wang1998semiclassical, sun1998semiclassical, miller2001semiclassical}. The accuracy of LSC-IVR depends on the particular choice of phase-space representation. We use the spin-mapping approach (spin-LSC)~\cite{runeson2019, runeson2020, miyazaki2023nonadiabatic}, which represents a spin-$S$ particle with $N = 2S+1$ levels as a point in the phase space of SU($N$)~\cite{weight2025ab, bossion2022non}.

For the $i$th spin, which has $N_i$ levels, spin mapping introduces $N_i$ pairs of Cartesian mapping variables, $\{X_{i,n}, P_{i,n}\}$, and collects them into the complex vector $\mathbf{z}_i$ with components $z_{i,n} = X_{i,n} + iP_{i,n}$. In the Stratonovich-Weyl W-representation~\cite{stratonovich1957, brif1998, brif1999, klimov2009group, tilma2012, tilma2016, runeson2020}, a single-spin operator $A$ maps onto the phase-space function $A_{\rm W}(\mathbf{z}_i) = \frac{1}{2}\left(\mathbf{z}_i^\dagger A \mathbf{z}_i - \gamma_i \mathrm{Tr}[A]\right)$, where $\gamma_i = \frac{2}{N_i}(\sqrt{N_i+1}-1)$ is the zero-point energy parameter. The traceless spin operators thus map onto the classical spin vector, $\mathcal{S}_i^k = \frac{1}{2}\mathbf{z}_i^\dagger S_i^k\mathbf{z}_i$, and mapping every spin operator in Eq.~\ref{eq:ham} gives the classical Hamiltonian, $H_{\rm W}$. Within the LSC approximation, expectation values become
\begin{align}
    \mathrm{Tr}[\rho(0)A(t)] \approx \int d\mathbf{z}(0) \rho_{\rm W}\left(\mathbf{z}(0)\right) A_{\rm W}\left(\mathbf{z}(t)\right),
    \label{eq:corr}
\end{align}
where $\mathbf{z}(t)$ evolves classically under $H_{\rm W}$ from $\mathbf{z}(0)$ and $\langle\cdot\rangle$ denotes the average over initial conditions sampled from $\rho_{\rm W}$.

Instead of sampling the full hypersphere, we use focused initial conditions~\cite{meyer1979, bonella2003semiclassical, bonella2005land, huo2012consistent}, which typically converge with an order of magnitude fewer trajectories. To focus the $i$th spin on the state $\ket{\phi_n}$ of an orthonormal basis $\{\ket{\phi_m}\}$, we set $\braket{\phi_m|\mathbf{z}_i} = r_m e^{i\varphi_m}$ with $r_m = \sqrt{2\delta_{m,n} + \gamma_i}$, and independent phases $\varphi_m$ drawn uniformly from $[0, 2\pi)$. Each trajectory then assigns unit population to $\ket{\phi_n}$ and zero population to all other basis states, and the coherences between them vanish upon averaging. To capture the multiplicatively separable initial condition for the sensor--bath system, as introduced in Sec.~\ref{sec:hamiltonian}, we sample randomly and uniformly from $I_i^z$ for every bath spin to capture its infinite-temperature distribution given by $\mathbb{1}/N_i$. For the sensor, the initial state $\ket{\psi_0} = (\ket{0}+\ket{1})/\sqrt{2}$ is not an eigenstate of $S_0^z$. Previous approaches to sample this type of initial condition span stochastic~\cite{breuer1997stochastic,sato2018coupled} and deterministic (rotation-based)~\cite{montoya2016approximate, runeson2019, amati2022quasiclassical} approaches. For the subspace of the cluster corresponding to the sensor's spin, we adopt a deterministic sampling. Because the initial state is the eigenstate for $S_0^x$ for a spin 1/2 sensor, we adopt the $S_0^x$ eigenbasis $\{\ket{\pm} = (\ket{0}\pm\ket{1})/\sqrt{2}\}$. One can follow the same protocol even for a higher-spin sensor by targeting specific orthogonal linear combinations that replace the $S_0^z$ eigenstates.

Averaging over trajectories then yields the sensor observables, $\bra{0}\rho_{\rm s}(t)\ket{1} \approx \frac{1}{2}\langle z_{0,0}(t) z_{0,1}^*(t)\rangle$ and $P_1(t) \approx \frac{1}{2}\langle |z_{0,1}(t)|^2 - \gamma_0\rangle$, where $\langle\cdot\rangle$ denotes the average over focused initial conditions.

\subsection{Classical propagation}

Along each trajectory, the mapping variables evolve under Hamilton's equations for the transformed Hamiltonian $H_{\rm W}$, which take a Schr\"odinger-like form,
\begin{equation}
    \dot{\mathbf{z}}_i = -i h_i(t) \mathbf{z}_i,
    \label{eq:eom}
\end{equation}
where $h_i(t)$ is given by
\begin{equation}
    h_i(t) = H_i + \sum_{j\neq i}\sum_{k,l\in\{x,y,z\}} S_i^k \mathbf{K}_{ij}^{kl}\,\mathcal{S}_j^l(t).
    \label{eq:hmf}
\end{equation}
Here, $H_i$ contains the single-spin terms of the $i$th spin. $\mathbf{K}_{ij}$ denotes the coupling tensor between spins $i$ and $j$, with $\mathbf{K}_{0j} = \mathbf{A}_j$ for the sensor ($i=0$) and $\mathbf{K}_{ij} = \mathbf{J}_{i,j}$ for bath spins. $\mathcal{S}_j^l(t)$ is the classical spin vector of the $j$th spin. Since $H_i$ enters Eq.~\ref{eq:eom} as an $N_i\times N_i$ matrix, this evolution treats single-spin terms, such as zero-field splittings and quadrupolar interactions, exactly and approximates only the spin-spin interactions. We integrate Eq.~\ref{eq:eom} with a fourth-order Runge-Kutta scheme and apply each pulse as an instantaneous rotation of the sensor mapping variables.

\begin{figure}
    \centering
    \includegraphics[width=\linewidth]{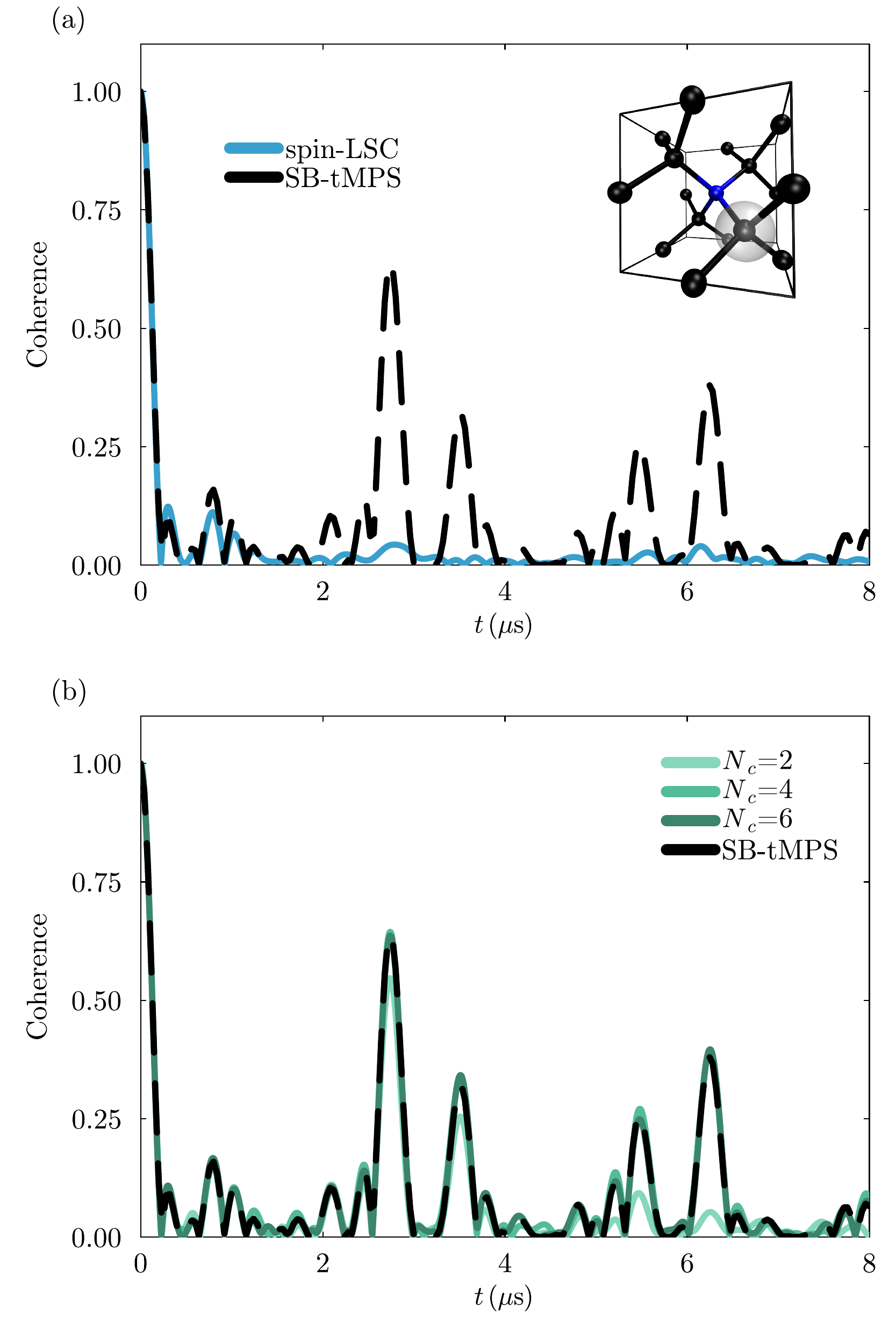}
    \caption{Coherence dynamics of an NV center in diamond with a bath of one $^{14}$N and 45 $^{13}$C nuclear spins at $B=100$~G. (a) Free induction decay obtained with spin-LSC (blue) and SB-tMPS (black dashed). Spin-LSC captures the initial dephasing but none of the subsequent coherence revivals. Inset: structure of the NV center in diamond, with nitrogen (blue) and the vacancy (white). (b) Cluster spin-LSC (CSLSC) dynamics with increasing bath spins in the cluster, $N_c = 2, 4, 6$ (light to dark green), compared with SB-tMPS (black dashed). CSLSC results converge systematically to the exact result, reaching quantitative agreement at $N_c=6$.}
    \label{fig:fig1}
\end{figure}

\subsection{Clustering}

\begin{figure*}[htb!]
    \centering
    \includegraphics[width=\linewidth]{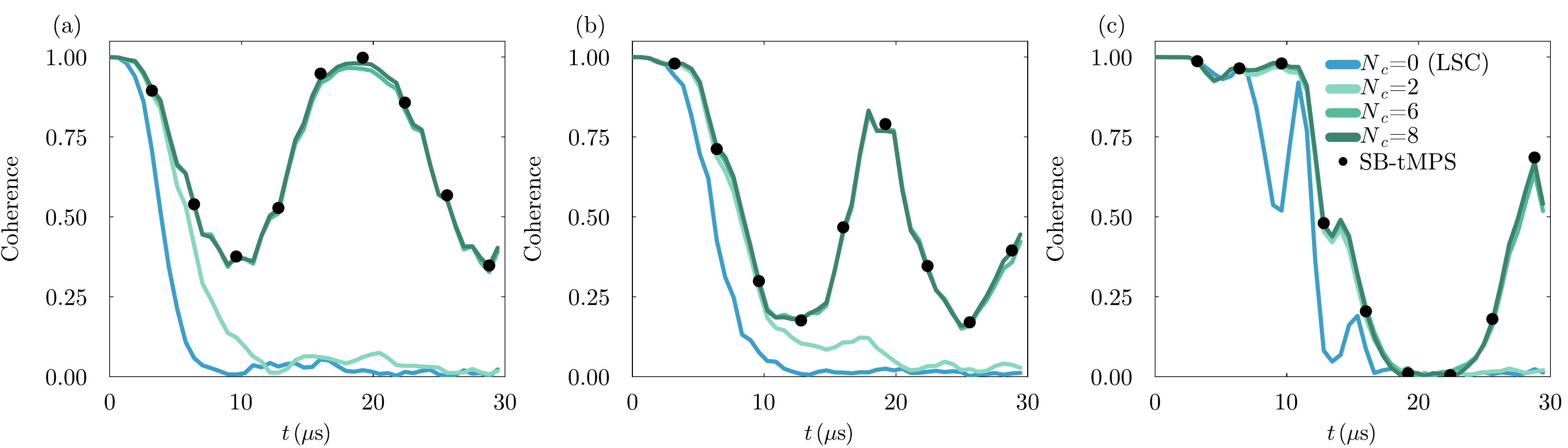}
    \caption{Coherence dynamics of the NV center under (a) spin echo, (b) CPMG-2, and (c) CPMG-4 pulse sequences. CSLSC results with $N_c = 0$ (spin-LSC, blue) and $N_c = 2, 6, 8$ (light to dark green), compared with SB-tMPS (black circles). Spin-LSC and small clusters miss coherence revivals, whereas CSLSC converges to the SB-tMPS result by $N_c=6$.}
    \label{fig:fig2}
\end{figure*}

The sensor-bath couplings in Eq.~\ref{eq:sbham} span a wide range of strengths and anisotropies. As discussed in \Sec{sec:hamiltonian}, most bath spins couple to the sensor predominantly through secular terms, which classical propagation describes faithfully, while the coherence revivals arise from a handful of nuclei with appreciable pseudosecular or quadrupolar couplings. Since each spin in Eq.~\ref{eq:hmf} interacts with other spins only through classical spin vectors, spin-LSC does not capture coherence revivals. Therefore, we group the sensor and $N_c$ bath spins into a cluster, $\mathcal{C}$, which we map as a single $D_{\mathcal{C}}$-level system, where $D_{\mathcal{C}} = \prod_{i\in\mathcal{C}} N_i$. One vector of mapping variables, $\mathbf{z}_{\mathcal{C}}$, with zero-point energy parameter $\gamma_{\mathcal{C}} = \frac{2}{D_{\mathcal{C}}}(\sqrt{D_{\mathcal{C}}+1}-1)$, describes the cluster. In each trajectory, we focus $\mathbf{z}_{\mathcal{C}}(0)$ on the product state $\ket{\psi_0}\otimes\ket{n}$, where $\ket{n}$ is a randomly chosen basis state of the bath spins in the cluster, consistent with our discussion of initial condition sampling in Sec.~\ref{subsec:phase-space}, and propagate it via
\begin{align}
    \dot{\mathbf{z}}_{\mathcal{C}} = -i\Big[H_{\mathcal{C}} + \sum_{i\in\mathcal{C}}\sum_{j\notin\mathcal{C}}\sum_{k,l\in\{x,y,z\}} S_i^k\mathbf{K}_{ij}^{kl}\mathcal{S}_j^l(t)\Big]\mathbf{z}_{\mathcal{C}}, \label{eq:cluster}
\end{align}
where $H_{\mathcal{C}}$ contains all single-spin and coupling terms within the cluster and $S_i^k$ acts on the $i$th member of the cluster. We treat couplings within the cluster quantum mechanically, whereas spins outside it enter through their classical spin vectors, as in Eq.~\ref{eq:hmf}. We obtain the sensor observables by tracing out the bath spins from $\mathbf{z}_{\mathcal{C}}$.

To choose which spins enter the cluster, we rank the bath spins by the magnitude of their pseudosecular coupling to the sensor, $|\mathbf{A}_i^{zx}|^2 + |\mathbf{A}_i^{zy}|^2$, and place the $N_c$ highest-ranked spins in the cluster. We place quadrupolar nuclei that are strongly coupled with the sensor directly in the cluster. When all pseudosecular terms vanish, only quadrupolar nuclei enter the cluster. Since $N_c = 0$ recovers spin-LSC and $N_c = K$ recovers the exact dynamics, $N_c$ can be systematically increased until convergence.

For large baths, storing $\mathbf{K}$ and multiplying by it at every time step becomes prohibitively expensive. The number of intra-bath couplings grows as $K^2$, which makes evaluating Eq.~\ref{eq:hmf} the dominant cost for large baths. Since dipolar couplings decay as $r^{-3}$, we retain only bath pairs with $\lVert\mathbf{J}_{i,j}\rVert \geq J_{\rm cut}$, which we identify with a cell-based neighbor list~\cite{allen2017computer}, and store $\mathbf{K}$ as a sparse matrix. Because the bath spins are fixed in space, we only need to build this list once. Each time step then costs $\mathcal{O}(D_{\mathcal{C}}^2 + {\rm nnz}(\mathbf{K}))$, where ${\rm nnz}(\mathbf{K})$ denotes the number of non-zero elements in $\mathbf{K}$. At fixed $N_c$ and $J_{\rm cut}$, the cost per trajectory grows linearly with the bath size. Since trajectories evolve independently, their evolution is trivially parallelizable.

\section{Results}

We now assess the performance of our CSLSC in predicting the decoherence dynamics of three paradigmatic spin platforms: a nitrogen-vacancy (NV) center in diamond, the [$^{167}$Er(hfac)$_4$]$^-$ molecular qubit solvated in deuterated \textit{o}-terphenyl (d-OTP), and an indium arsenide quantum dot. Beyond their individual promise for nanoscale sensing~\cite{Lovchinsky2016, Aslam2023}, telecom-compatible spin-photon interfaces~\cite{Weiss2025}, and solid-state quantum memories~\cite{gangloff2019quantum}, these platforms span a wide range of the parameter space of Eq.~\ref{eq:ham}, increasingly push the limits of existing methods for spin dynamics, and challenge our method in different ways. The NV center hosts a sparse and weakly interacting $^{13}$C bath that places its dynamics close to the pure-dephasing limit. The [Er(hfac)$_4$]$^-$ complex, on the other hand, lies well beyond the pure-dephasing limit, and the strongly anisotropic $g$-tensor of its Kramers doublet couples the electronic spin to the $I=7/2$ $^{167}$Er nucleus of the metal center and to the ligand $^{19}$F and $^1$H nuclei, which imprint pronounced modulations on its echo dynamics~\cite{Weiss2025}. Embedding the complex in d-OTP adds $\sim$200 solvent deuterium nuclei with $I=1$, rendering the solvated system high-dimensional and difficult to treat beyond a direct semiclassical level. Finally, the electron spin in an InAs quantum dot interacts with $10^4$--$10^5$ high-spin nuclei ($I=3/2$ and $9/2$) whose quadrupolar couplings drive distinct stages of decoherence that unfold from nanoseconds to microseconds~\cite{Bulutay2012,bechtold2015three}. Resolving this behavior requires the long-time dynamics of a large, interacting, high-spin bath---a regime far beyond the reach of numerically exact methods, including our SB-tMPS~\cite{Li2026}. Hence, we benchmark our method against SB-tMPS for the NV center and the isolated erbium complex (SI~Sec.~II), and then expand it to regimes where no numerically exact benchmark is available, specifically, the solvated erbium complex and the quantum dot.

The NV center is a point defect in diamond formed by a substitutional nitrogen adjacent to a lattice vacancy, whose negatively charged state hosts an $S=1$ electronic ground state. Its coherence times approach a millisecond even at room temperature~\cite{stanwix2010coherence}, which has enabled nanoscale magnetometry and electrometry~\cite{maze2008nanoscale, dolde2011electric, liu2019coherent, DegenRevModPhys2017}, nuclear magnetic resonance spectroscopy of single proteins~\cite{Lovchinsky2016, Aslam2023}, and photonic quantum networks~\cite{nemoto2016photonic}. Here, we couple the electronic spin of the NV center to a bath of 46 nuclear spins: the quadrupolar $I=1$ $^{14}$N nucleus of the defect, and 45 $^{13}$C nuclei with $I=1/2$ in the surrounding lattice under an applied magnetic field of $B=100$~G. The intermediate size of the system and weak intrabath interactions make this system an ideal test case for our method since it remains amenable to numerically exact SB-tMPS benchmarks while testing the ability of our cluster to capture the coherent modulations generated by the quadrupolar $^{14}$N nucleus and the few $^{13}$C nuclei with appreciable pseudosecular hyperfine couplings. We obtain the hyperfine tensors of the nuclei closest to the defect and the electric field gradient at the $^{14}$N site, which sets its quadrupolar coupling, from density functional theory calculations with \texttt{Quantum ESPRESSO}~\cite{pickard2001all, petrilli1998electric, charpentier2011paw, giannozzi2009quantum, giannozzi2017advanced} (see SI Sec.~1).

\begin{figure*}[htb!]
    \centering
    \includegraphics[width=\linewidth]{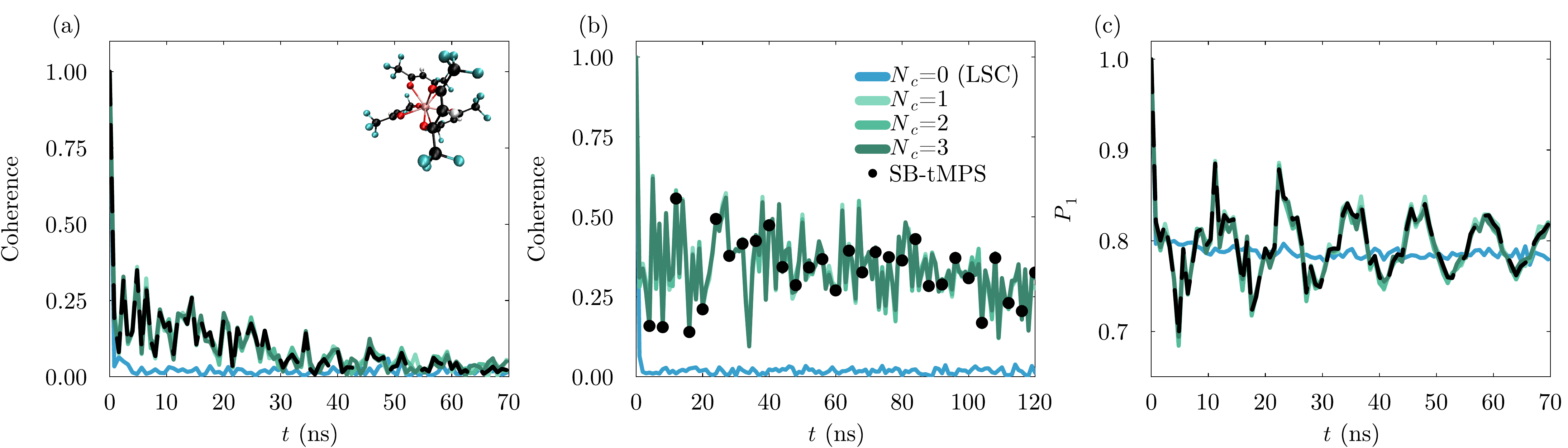}
    \caption{Dynamics of the isolated [Er(hfac)$_4$]$^-$ complex with a bath of the $^{167}$Er, 24 $^{19}$F, and 4 $^1$H nuclear spins at $B=240$~G. (a) Free induction decay and (b) spin echo coherence dynamics, and (c) population dynamics of the sensor initialized in $\ket{1}$, obtained with CSLSC with $N_c = 0$ (spin-LSC, blue) and $N_c = 1$, 2, and 3 (light to dark green), compared with SB-tMPS (black). Including only the $^{167}$Er nucleus in the cluster ($N_c=1$) recovers the SB-tMPS dynamics. Inset: structure of the [Er(hfac)$_4$]$^-$ complex.}
    \label{fig:fig3}
\end{figure*}

Figure~\ref{fig:fig1}(a) compares the free induction decay of the NV center coherence obtained from spin-LSC and SB-tMPS. Both methods agree on the initial decay, in which the coherence falls below 0.1 within $\sim0.2\mu$s, and spin-LSC even recovers the small-amplitude oscillations that follow up to $t\approx1~\mu$s. Beyond this timescale, however, the SB-tMPS dynamics exhibit pronounced coherence revivals, the largest of which reaches $C(t)\approx0.64$ at $t\approx2.7~\mu$s, followed by further revivals at $t\approx3.5$, $5.5$, and $6.2~\mu$s. Spin-LSC captures none of these revivals and instead predicts a coherence that remains below $0.05$ for the rest of the simulation. Under classical propagation, spin-LSC fails to build the sensor-bath entanglement that underlies ESEEM-like revivals. Figure~\ref{fig:fig1}(b) shows that our CSLSC systematically recovers these revivals as we increase $N_c$. With two bath spins in the cluster, $N_c=2$ reproduces the position and amplitude of the dominant revival at $t\approx2.7~\mu$s, but underestimates subsequent revivals, most noticeably at $6.2~\mu$s, where it predicts amplitudes below the SB-tMPS result. Increasing the cluster size to $N_c=4$ recovers all major revivals, with residual deviations in the amplitudes. At $N_c=6$, the CSLSC and SB-tMPS results agree quantitatively over the entire simulation time. 

The dynamics under CPMG pulse sequences further expose the consequences of missing these revivals. Figure~\ref{fig:fig2} compares the coherence dynamics of the NV center under spin echo (a), CPMG-2 (b), and CPMG-4 (c) sequences obtained with CSLSC at increasing $N_c$ and with SB-tMPS. Refocusing pulses cancel the dephasing caused by the quasi-static Overhauser field, and the remaining coherence dynamics are governed by the sensor-bath entanglement that semiclassical approximations normally fail to describe. As a result, spin-LSC fails to capture the entire coherence decay. It predicts a coherence that decays too quickly and never revives. Small clusters partially correct the initial decay, but $N_c=2$ still misses the pronounced revivals that dominate the echo and CPMG signals. In contrast, increasing the cluster size to $N_c=6$ recovers the full SB-tMPS dynamics for all three pulse sequences, and further enlarging the cluster to $N_c=8$ leaves the results unchanged, confirming convergence with cluster size. 

This accuracy comes at a fraction of the computational cost of using SB-tMPS. Each point of a CPMG signal corresponds to a different pulse spacing, $\tau = t/2n$, and thus requires a simulation up to time $t$. SB-tMPS calculations for the NV center pulse-sequence dynamics in Fig.~\ref{fig:fig2} require $\sim 33$ hours with 5 NVIDIA A100 GPUs running in parallel to obtain different data points, despite yielding half the number of data points as CSLSC. In contrast, CSLSC, parallelized on a CPU, yields the converged $N_c=6$ dynamics in $\sim$8 minutes.

The systematic convergence of CSLSC with $N_c$ supports the physical picture of \Sec{sec:hamiltonian}: the coherence revivals of the NV center are dominated by a small number of nuclei with appreciable pseudosecular or quadrupolar couplings. Classical propagation accurately describes the remaining $\sim40$ nuclei, which couple predominantly through secular terms. Moreover, each enlargement of the cluster resolves revivals at progressively later times, suggesting that nuclei with weaker pseudosecular couplings, which rank lower in our selection hierarchy, imprint their signatures on the coherence at correspondingly longer times. Achieving numerically exact agreement requires treating only 6 of the 46 bath spins within the cluster, such that the cluster spans a Hilbert space of at most a few hundred states, while the remaining bath spins are propagated at the cost of independent single-spin dynamics. Unlike factorization approaches, whose results need not improve with increasing expansion order~\cite{Li2026, ryan2026challenge}, enlarging the cluster in our method systematically approaches the exact dynamics. The agreement between successive cluster sizes therefore allows for an internal convergence criterion that does not rely on an exact benchmark. We exploit this for systems that lie beyond the reach of SB-tMPS.

% Result 3 : Erbium complex FID, SE and populations (3-panel)   

\begin{figure*}
    \centering
    \includegraphics[width=\linewidth]{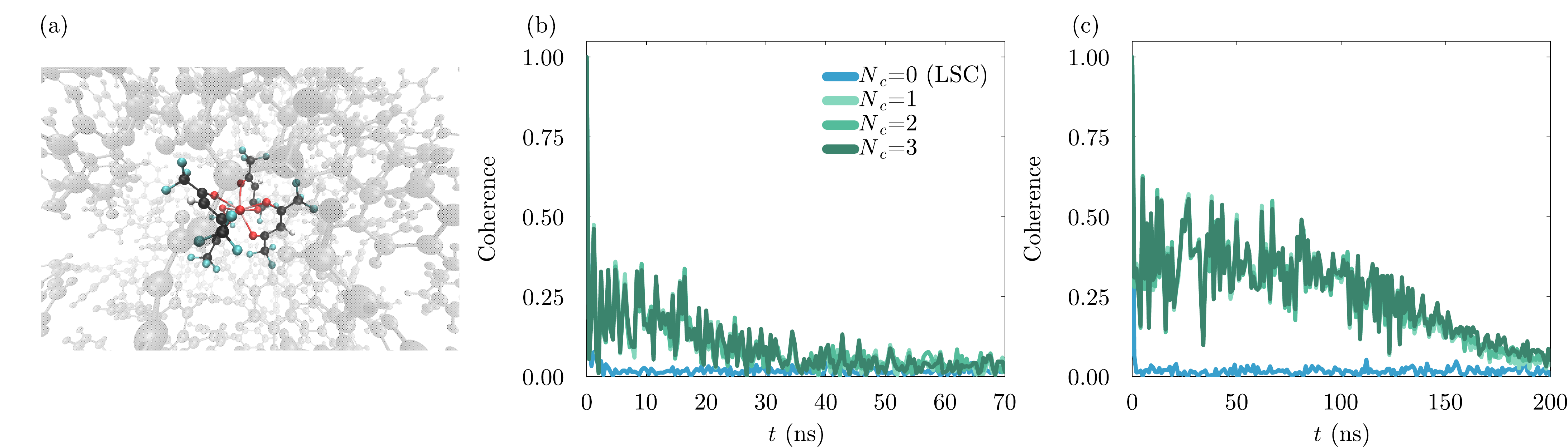}
    \caption{Coherence dynamics of [Er(hfac)$_4$]$^-$ solvated in d-OTP. (a) Schematic of the complex embedded in the d-OTP solvent shell containing 200 deuterium nuclei. (b) Free induction decay and (c) spin echo coherence dynamics obtained with CSLSC with $N_c = 0$ (spin-LSC, blue) and $N_c = 1$, 2, and 3 (light to dark green). The results converge at $N_c=1$.}
    \label{fig:fig4}
\end{figure*}

We next turn to molecular spin qubits, whose synthetic tunability offers a route to modular design in quantum information. Among these, lanthanide complexes have large and anisotropic magnetic moments with optical transitions between shielded $4f$ orbitals~\cite{gaita2019molecular, kumar2021optical}, and have attracted growing theoretical interest~\cite{briganti2021complete, sansone2026raman}. In the [Er(hfac)$_4$]$^-$ complex (shown in the inset of Fig.~\ref{fig:fig3}(a)), the ground Kramers doublet of Er$^{3+}$ behaves as an effective $S=1/2$ spin with a strongly anisotropic $g$-tensor and couples to optical transitions at telecom wavelengths, enabling optical spin initialization and readout that resolves both spin states and magnetically inequivalent sites~\cite{Weiss2025}. To enable a direct comparison with SB-tMPS, we first consider the isolated complex, whose bath comprises 29 nuclear spins: the quadrupolar $I=7/2$ $^{167}$Er nucleus, and the 24 $^{19}$F and 4 $^1$H nuclei from the hfac ligands. We apply a magnetic field of $B=240$~G, consistent with low-field spin-resolved measurements~\cite{Weiss2025}, and retain all components of the hyperfine tensors, such that the dynamics lie beyond the pure-dephasing limit. We obtain the $g$-tensor and hyperfine tensors of the ground Kramers doublet and the quadrupolar coupling of the Er nucleus from \textit{ab initio} calculations with ~\texttt{ORCA}~\cite{neese2020orca, chibotaru2012ab, reiher2012relativistic, chilton2025ab} (see SI Sec.~1).

Figure~\ref{fig:fig3} compares the coherence and population dynamics of the isolated [Er(hfac)$_4$]$^-$ complex obtained with CSLSC and SB-tMPS. Spin-LSC only captures the initial decay of the free induction decay (Fig.~\ref{fig:fig3}(a)) and misses the subsequent revivals, as in the case of the NV center. Again, this becomes more pronounced under a spin echo (Fig.~\ref{fig:fig3}(b)), where SB-tMPS coherence remains strongly modulated while the spin-LSC coherence vanishes almost immediately. Including only the $^{167}$Er nucleus in the cluster ($N_c=1$) suffices to recover most of the revivals for both measurements, and further enlarging the cluster changes the dynamics negligibly. 

In molecular qubits, anisotropic hyperfine couplings can also drive population relaxation on timescales comparable to dephasing, opening an additional decoherence channel~\cite{ficheux2018dynamics}. Figure~\ref{fig:fig3}(c) compares the population dynamics of the sensor initialized in $\lvert 1 \rangle$ obtained using spin-LSC, CSLSC, and SB-tMPS. Within the first nanosecond, all methods capture the rapid relaxation of $P_1(t)$ to $\approx0.8$. Beyond this point, the SB-tMPS dynamics exhibit persistent oscillations about this value, whereas spin-LSC predicts a nearly constant population that misses these oscillations entirely. In contrast, CSLSC reproduces these oscillations from $N_c=1$ onward and matches the SB-tMPS result, showing that CSLSC provides an accurate prediction for spin dynamics beyond the pure-dephasing limit.

% Result 4 : Erbium complex with solvent (3-panel)

Realistically, molecular qubits are rarely studied in isolation and are typically dissolved in frozen solutions or embedded in glassy matrices whose nuclear spins form an extended bath~\cite{zadrozny2015millisecond, yu2016long, Kveder2019JCP, Canarie2020JPCL}. To capture this environment, we embed the complex in a d-OTP solvent shell, adding 200 deuterium ($I=1$) nuclei to the bath (see Fig.~\ref{fig:fig4}(a)). Together with the 29 nuclear spins of the complex, the resulting Hilbert space has a dimension of $\sim$$10^{105}$, placing the solvated system far beyond the reach of numerically exact methods, including SB-tMPS, and making a scalable semiclassical treatment essential.

In Fig.~\ref{fig:fig4}, we show the coherence dynamics of the solvated complex under free induction decay (b) and spin echo (c). As in the isolated complex, the cluster spin-LSC results are converged at $N_c=1$, providing the internal convergence criterion established for the NV center in the absence of an exact benchmark. Our cluster spin-LSC thus extends accurate simulations of molecular spin qubits to realistic solvation environments that are inaccessible to numerically exact methods.
 
% Result 4 : Quantum dot (absence of benchmark)

\begin{figure}[htb!]
    \centering
    \includegraphics[width=\linewidth]{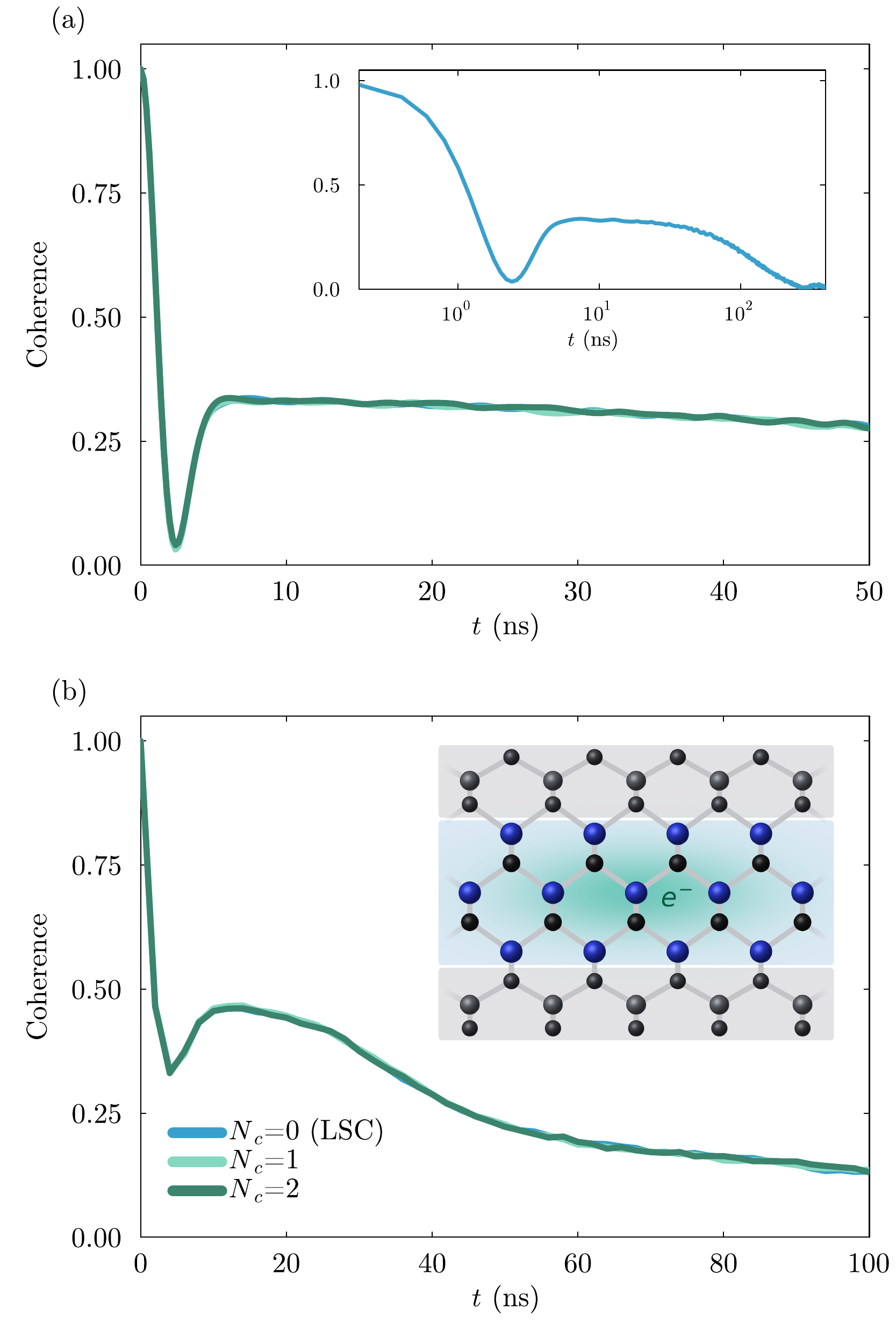}
    \caption{Coherence dynamics of an electron spin in an InAs quantum dot with a bath of 35,000 nuclear spins at zero magnetic field. (a) Free induction decay and (b) spin echo coherence dynamics obtained with cluster spin-LSC with $N_c = 0$ (spin-LSC, blue) and $N_c = 1, 2$ (light to dark green). The results are insensitive to cluster size. Inset of (a): long-time free induction decay obtained with spin-LSC on a logarithmic time axis. Inset of (b): schematic of the self-assembled InAs quantum dot with the confined electron delocalized within the dot.}
    \label{fig:fig5}
\end{figure}

We now consider an electron spin confined in a self-assembled InAs quantum dot, a semiconductor platform that combines bright, coherent optical transitions with an isolated mesoscopic ensemble of nuclear spins that can serve as a local quantum memory~\cite{gangloff2019quantum}. To construct spin Hamiltonian parameters for the dot, we start from a zinc-blende InAs cell containing 35,000 nuclear spins in the absence of an external magnetic field: 17,534 In nuclei, with natural isotopic abundances of $^{113}$In and $^{115}$In (both with $I=9/2$), and 17,466 $^{75}$As nuclei ($I=3/2$). The electron occupies an $s$-like conduction-band state that is delocalized over the entire cell such that it couples to every nucleus through the isotropic Fermi contact interaction, with a strength set by the electron density at the nuclear site. We detail the electronic wavefunction and the resulting hyperfine, quadrupolar, and bath-bath couplings in SI Sec.~1. The resulting Hilbert space has a dimension of $\sim10^{28,050}$, many orders of magnitude beyond that of the solvated erbium complex, and thus requires a semiclassical approximation~\cite{Merkulov2002, Khaetskii2002, Coish2004, Witzel2008, AlHassanieh2006, Chen2007, cywinski2009pure}. Existing semiclassical treatments of quantum dots typically neglect intrabath interactions and either replace the high-spin nuclei with effective $I=1/2$ spins and coarse-grain the bath into a reduced set of collective spins, or mimic quadrupolar couplings with random static fields~\cite{Stanek2014, Fauseweh2017, Rohrig2018, Sinitsyn2012, bechtold2015three}. On the other hand, our SU($N$) mapping retains every nucleus with its physical spin~\cite{runeson2020} and thus evolves quadrupolar interactions exactly. Further, by systematically adding bath spins to the cluster, we can now directly assess whether a semiclassical description of the electron and its bath is sufficient.

Figure~\ref{fig:fig5}(a) shows the free induction decay of the quantum dot electron spin. Within the first few nanoseconds, the coherence drops to a minimum and recovers to a plateau at $\approx1/3$ of its initial value. This behavior arises from the Overhauser field being effectively frozen and randomly oriented~\cite{Merkulov2002, bechtold2015three}. The inset of Fig.~\ref{fig:fig5}(a) shows the long-time dynamics of spin-LSC. The plateau persists for tens of nanoseconds before slowly decaying as the nuclear spin states change and the Overhauser field starts to relax. Under spin echo (Fig.~\ref{fig:fig5}(b)), the refocusing pulse partially removes the dephasing from the quasi-static Overhauser field, and the subsequent decay of the coherence reflects the fluctuations of the nuclear spin states. Unlike the NV center and the erbium complex, the dynamics remain unchanged upon adding bath spins to the cluster for both measurements. Hence, the large number of weakly and isotropically coupled nuclei in the quantum dot renders a semiclassical description of the electron and its bath sufficient, which our cluster spin-LSC allows us to establish directly in the absence of a numerically exact benchmark.

\section{Conclusion}

We have thus introduced CSLSC, a semiclassical approach to simulate the coherence and population dynamics of quantum sensors subject to an arbitrary number of dynamical decoupling pulses. CSLSC is a systematically convergent semiclassical method that recovers numerically exact dynamics in the limit where all spins are incorporated into the cluster. While doing this for anything but small systems ($\sim$14 spins) is infeasible, we observe converged dynamics that recover exact benchmarks for even small cluster sizes. The reason for this advantage lies in the hierarchy of interactions in many-spin systems that our method exploits: while most bath spins couple to the sensor through secular terms that semiclassics describes faithfully, only a few possess appreciable pseudosecular or quadrupolar couplings that require full quantum mechanical treatment. Importantly, our design of CSLSC is modular, meaning that one can replace the LSC approximation with any method in the ladder of quasiclassical~\cite{sun1997mixed, sun1998semiclassical, wang1998semiclassical, thoss2001generalized, wang2001systematic, huo2011PLDM, huo2012consistent, li2012cartesian, levy2019complete, pfalzgraff2019efficient} or semiclassical~\cite{walton1996new, thompson1999influence, makri1998semiclassical, ceotto2009multiple, thoss2004semiclassical, laskowski2026improve} approximations, as well as replace the exact diagonalization treatment of the cluster with a different numerically exact~\cite{wang2003multilayer, bartsch2009dynamical, lindoy2018simple, ren2022time, lindoy2025pyttn} or high-accuracy~\cite{carleo2017solving, karlsson2018generalized} method. Even at the level of the LSC approximation for the semiclassical treatment, CSLSC stays accurate across a wide swath of parameter regimes and reproduces the numerically exact SB-tMPS dynamics at a fraction of the cost. Unlike factorization approaches, CSLSC does not require a division of subcluster contributions and hence converges systematically with cluster size, providing an internal convergence check that extends accurate simulations to spin systems far beyond the reach of numerically exact methods.

Since CSLSC scales to realistic spin environments, it offers a path to understanding how the environment of a real sensor shapes its noise spectrum~\cite{alvarez2011measuring, cywinski2014dynamical, Szankowski2017JP, sung2019non}. Further, since LSC-IVR techniques accurately describe the dynamics of spin-boson models~\cite{sun1998semiclassical, wang1998semiclassical, runeson2019, runeson2020}, we anticipate that CSLSC should be able to treat systems where both spin-spin and spin-phonon interactions span commensurate energy scales. We anticipate that CSLSC, in conjunction with noise spectroscopy protocols~\cite{viola1999dynamical, alvarez2011measuring, norris2016qubit, paz2017multiqubit, vezvaee2024fourier, shitara2024fast}, can connect measured spectra to the density, depth, and arrangement of its environment. It is thus well suited to guide the design of nanoscale NMR technologies based on a single sensor, which exploit the entanglement between the sensor and individual nuclear spins~\cite{Zhao2012, Mamin2013, Staudacher2013, Lovchinsky2016, Aslam2017, Abobeih2019}.

\section*{Acknowledgements}

The authors acknowledge funding from the National Science Foundation (Grant No.~2412615). A.M.C.~also acknowledges the support from a David and Lucile Packard Fellowship for Science and Engineering. This work was supported by an Alfred P. Sloan Research Fellowship and the Camille and Henry Dreyfus Foundation through a Camille Dreyfus Teacher-Scholar Award. This work utilized the Alpine high-performance computing resource at the University of Colorado Boulder. Alpine is jointly funded by the University of Colorado Boulder, the University of Colorado Anschutz, Colorado State University, and the National Science Foundation (Award No. 2201538). This work used the Anvil cluster at Purdue University through allocation CHE250160 from the Advanced Cyberinfrastructure Coordination Ecosystem: Services and Support (ACCESS) program~\cite{Song:2022, Brown:2021}.

\bibliography{main}

\pagebreak

\end{document}

% \clearpage

\setcounter{section}{0}
\setcounter{equation}{0}
\setcounter{figure}{0}
\setcounter{table}{0}
\setcounter{page}{1}

\renewcommand{\theequation}{S\arabic{equation}}
\renewcommand{\thefigure}{S\arabic{figure}}
\renewcommand{\thepage}{S\arabic{page}}
\renewcommand{\bibnumfmt}[1]{$^{\mathrm{S#1}}$}
\renewcommand{\citenumfont}[1]{S#1}

\title{Supplementary Information for \\``Systematically improvable semiclassical clustering for accurate and efficient quantum dynamics in large spin ensembles''}

\author{Pranay Venkatesh}
\affiliation{Department of Chemistry, University of Colorado Boulder, Colorado 80309, USA}
\author{Andr\'es Montoya-Castillo}
\homepage{Andres.MontoyaCastillo@colorado.edu}
\affiliation{Department of Chemistry, University of Colorado Boulder, Colorado 80309, USA}

\maketitle

\renewcommand{\theequation}{S\arabic{equation}}
\renewcommand{\thefigure}{S\arabic{figure}}
\renewcommand{\thepage}{S\arabic{page}}
\renewcommand{\bibnumfmt}[1]{$^{\mathrm{S#1}}$}
\renewcommand{\citenumfont}[1]{S#1}

\onecolumngrid

\tableofcontents

\section{Hamiltonian Parametrization}

Realistic simulations of the spin dynamics in physical systems rely on the accurate parametrization of the underlying Hamiltonian described in Sec.~II of the main text. Specifically, we require accurate $g-$tensors of the sensor and bath spins, the hyperfine coupling tensors between the sensor and the bath spins ($\mathbf{A}$), quadrupolar tensors of the bath spins ($\mathbf{P}$), and the coupling tensors among the bath spins ($\mathbf{J}$).

For all systems considered in the main text, we compute bath-bath interaction tensors ${\bf J}_{i, j}$ from a point dipole approximation~\cite{abragam1961principles, schweiger2001principles} using a coupling tensor of the form:
\begin{align}
    \label{eq:pd_Jtensor}
    {\bf J}_{i,j}^{kl} = - \frac{\mu_0}{4\pi} \gamma_i \gamma_j \hbar \left[ \frac{3 {\bf r}_{i,j}^k  {\bf r}_{i,j}^l - | {\bf r}_{i,j}|^2 \delta_{kl}}{|{\bf r}_{i,j}|^5}  \right]
\end{align}
for bath spins $i$ and $j$ separated by ${\bf r}_{i,j}$. This approximation holds since the nuclear spins in the bath are fixed in place~\cite{Krogmeier2026}. Calculation of the $g-$tensors, $\mathbf{A}$, and $\mathbf{P}$ requires greater care, depending on the system. We explain how we calculated these terms for each system of the main text below. 

\subsection{NV center}

In the NV center, the sensor is an $S=1$ electronic spin that couples to 45 $I=1/2$ $^{13}$C nuclei, placed at random lattice sites with natural abundance (1.1\%), and to the quadrupolar $^{14}$N nucleus of the defect. We take the $z$ axis along the N-V axis and apply $B=100$~G along it. The sensor's dynamics lie in the pure-dephasing limit, and only the secular and pseudosecular components enter the Hamiltonian.

For the nuclei within 10~\AA\ of the vacancy, we obtain the hyperfine tensors from the electron spin density, $\rho_s({\bf r})$, computed with density functional theory,
\begin{align}
    \label{eq:atensor_est}
    {\bf A}_{i}^{kl} = - \frac{\mu_0}{4\pi} \frac{\gamma_s \gamma_i \hbar}{2S} \left( \frac{8 \pi}{3} \rho_s({\bf r}_i) \delta_{kl} + \int d{\bf r}\, \rho_s({\bf r}+{\bf r}_i) \left[ \frac{3 {\bf r}^k {\bf r}^l - | {\bf r}|^2 \delta_{kl}}{|{\bf r}|^5} \right] \right),
\end{align}
where ${\bf r}_i$ is the position of the $i$th nucleus. The first term is the Fermi contact interaction and the second is the spin-dipolar interaction. We compute $\rho_s$ with \texttt{Quantum ESPRESSO}~\cite{giannozzi2009quantum, giannozzi2017advanced} for the negatively charged NV center in a $4\times4\times4$ diamond supercell containing 511 atoms, using the PBE functional and GIPAW pseudopotentials~\cite{pickard2001all}. This applies to the $^{14}$N and the three $^{13}$C nuclei closest to the vacancy. For all other nuclei, we use the point-dipole limit of Eq.~\ref{eq:atensor_est}, in which the spin density is concentrated at the vacancy,
\begin{align}
    \label{eq:pd_Atensor}
    {\bf A}_{i}^{kl} = - \frac{\mu_0}{4\pi} \gamma_s \gamma_i \hbar \left[ \frac{3 {\bf r}_{i}^k {\bf r}_{i}^l - | {\bf r}_{i}|^2 \delta_{kl}}{|{\bf r}_{i}|^5} \right].
\end{align}
We obtain the quadrupolar tensor of the $I=1$ $^{14}$N nucleus from the electric field gradient tensor at the nitrogen site using GIPAW~\cite{pickard2001all, petrilli1998electric, charpentier2011paw}.

\subsection{Erbium complex}

The sensor corresponds to the ground Kramers doublet of Er$^{3+}$, which behaves as an effective spin $S=1/2$~\cite{Weiss2025}. Its $g-$tensor is highly anisotropic, with the Zeeman interaction given by $H_Z = \mu_B \mathbf{B}\cdot\mathbf{g}\cdot\mathbf{S}_0$. The sensor thus precesses about the axis $\hat{\mathbf{n}} \propto \mathbf{g}^{T}\mathbf{B}$, at frequency $\omega_s = \mu_B|\mathbf{g}^{T}\mathbf{B}|/\hbar$.

We obtain $\mathbf{g}$, the hyperfine tensors, and the electric field gradients from a single \texttt{ORCA}~\cite{neese2020orca} calculation on the [Er(hfac)$_4$]$^-$ complex. We use CASSCF(11,7) over the $4f^{11}$ configuration, followed by quasi-degenerate perturbation theory for spin-orbit coupling, with the second-order Douglas-Kroll-Hess Hamiltonian~\cite{reiher2012relativistic}. We project $\mathbf{g}$ and the hyperfine tensors of the $^{167}$Er ($I=7/2$), 24 $^{19}$F, and 4 $^1$H nuclei onto the ground Kramers doublet, in the same pseudospin basis~\cite{chibotaru2012ab}. We use point dipole couplings for the nearest 200 deuterium nuclei of the d-OTP solvent, and neglect their quadrupolar couplings. We compute the quadrupolar tensor of the $I=7/2$ $^{167}$Er nucleus from the electric field gradient at the Er nucleus using \texttt{ORCA}~\cite{chilton2025ab}.

\subsection{Quantum dot}

Since the self-assembled quantum dot is a mesoscopic object whose electron wavefunction extends over $\sim 10^4$--$10^5$ lattice sites, outside the reach of a first principles electronic structure theory calculation, we describe the confined electron with a model envelope wavefunction within the effective-mass approximation~\cite{Coish2004, Witzel2008}. We build the bath from a zinc-blende InAs lattice (lattice constant $a = 6.058$~\AA) with the growth axis [111] along $z$, with natural abundances of $^{113}$In and $^{115}$In ($I=9/2$) and $^{75}$As ($I=3/2$). The electron in the $s-$like conduction band has a normalized anisotropic Gaussian envelope,
\begin{align}
    \label{eq:qd_envelope}
    |\psi({\bf r})|^2 = \frac{1}{\pi^{3/2} a_\perp^2 a_z} \exp\left( -\frac{x^2+y^2}{a_\perp^2} - \frac{z^2}{a_z^2} \right),
\end{align}
with lateral and vertical widths $a_\perp = 100$~\AA\ and $a_z = 30$~\AA, centered on the dot. We keep the 35,000 nuclei with the largest $|\psi({\bf r}_i)|^2$.

Each nucleus couples to the electron through the isotropic Fermi contact interaction,
\begin{align}
    \label{eq:qd_Atensor}
    {\bf A}_i^{kl} = \frac{\mathcal{A}_i}{\hbar}\, v_0 |\psi({\bf r}_i)|^2 \delta_{kl},
\end{align}
where $v_0 = a^3/4$ is the volume per InAs formula unit and $\mathcal{A}_i$ is the bulk hyperfine constant of the nuclear species ($\mathcal{A}_{\rm In} = 56~\mu$eV and $\mathcal{A}_{\rm As} = 46~\mu$eV)~\cite{cywinski2009pure}. Since every component of $\mathbf{A}_i$ is retained at zero field, the dynamics lie beyond the pure-dephasing limit. This form follows the standard treatment of the electron-nuclear hyperfine coupling in quantum dots~\cite{Coish2004}.

Every bath spin has $I>1/2$ and hence a quadrupolar interaction, which in self-assembled dots arises from the electric field gradients induced by lattice strain. We use a uniform-strain model~\cite{Bulutay2012}, in which every quadrupolar tensor is axially symmetric about the growth axis.

The bath-bath couplings follow Eq.~\ref{eq:pd_Jtensor}. Since the bath is dense, we retain only the pairs whose coupling exceeds a threshold of $J_{\rm cut} = 28$~Hz, where dynamics are converged. 

\section{Details of SB-tMPS calculations}

For the benchmark SB-tMPS~\cite{Li2026} calculations in Figs.~1, 2, and 3 of the main text, we represent the many-spin density matrix as a vectorized matrix product state (MPS) and propagate the initial state with the time-dependent variational principle, which fixes the MPS bond dimension throughout the evolution. We employ a converged bond dimension of $\chi = 100$ for all simulations. In constructing the matrix product operators, we impose a hierarchy of singular value decomposition (SVD) truncation thresholds informed by the relative strengths of the Hamiltonian couplings: the sensor, bath, and sensor-bath coupling terms are retained at a standard high-precision threshold, while the intrabath terms are truncated with $r_{\rm tr} = 10^{-4}$. This aggressive intrabath truncation holds for systems where sensor-bath and intrabath interactions are energetically separated~\cite{Li2026}.

%\section{References}
\clearpage
\bibliography{main}